\documentclass[pra,reprint,superscriptaddress,bibnotes,floatfix]{revtex4-1}  

\usepackage[T1]{fontenc}
\usepackage[utf8]{inputenc} 
\usepackage[australian]{babel}
\usepackage[a4paper,centering,hmargin=1.75cm,vmargin=2cm]{geometry} 
\usepackage{amsmath,amssymb,graphicx,bm,microtype} 
\usepackage[dvipsnames]{xcolor}
\usepackage[colorlinks,allcolors=blue!50!black]{hyperref}
\usepackage[all]{hypcap}
\usepackage{cleveref}
\usepackage{enumitem}
\usepackage{titlesec}
\titleformat{\paragraph}[runin]{\bfseries}{}{0pt}{}[.]
\usepackage{braket}
\usepackage{siunitx}

\newcommand{\abs}[1]{\lvert #1 \rvert}
\renewcommand{\vec}[1]{\bm{\mathrm{#1}}}
\let\Re\relax \DeclareMathOperator{\Re}{Re}

\begin{document}

\title{Delocalised kinetic Monte Carlo for simulating delocalisation-enhanced charge and exciton transport in disordered materials}

\author{Daniel Balzer}
\affiliation{School of Chemistry and University of Sydney Nano Institute, University of Sydney, NSW 2006, Australia}

\author{Thijs J.A.M. Smolders}
\affiliation{School of Chemistry and University of Sydney Nano Institute, University of Sydney, NSW 2006, Australia}
\affiliation{Institute for Molecules and Materials, Radboud University, 6525 AJ Nijmegen, The Netherlands}

\author{David Blyth}
\affiliation{School of Mathematics and Physics, University of Queensland, St.\ Lucia, QLD 4072, Australia}

\author{Samantha N. Hood}
\affiliation{School of Mathematics and Physics, University of Queensland, St.\ Lucia, QLD 4072, Australia}

\author{Ivan Kassal}
\email[Email: ]{ivan.kassal@sydney.edu.au}
\affiliation{School of Chemistry and University of Sydney Nano Institute, University of Sydney, NSW 2006, Australia}

\begin{abstract}
Charge transport is well understood in both highly ordered materials (band conduction) or highly disordered ones (hopping conduction). In moderately disordered materials—including many organic semiconductors—the approximations valid in either extreme break down, making it difficult to accurately model the conduction. In particular, describing wavefunction delocalisation requires a quantum treatment, which is difficult in disordered materials that lack periodicity. Here, we present the first three-dimensional model of partially delocalised charge and exciton transport in materials in the intermediate disorder regime. Our approach is based on polaron-transformed Redfield theory, but overcomes several computational roadblocks by mapping the quantum-mechanical techniques onto kinetic Monte Carlo. Our theory, delocalised kinetic Monte Carlo (dKMC), shows that the fundamental physics of transport in moderately disordered materials is that of charges hopping between partially delocalised electronic states. Our results reveal why standard kinetic Monte Carlo can dramatically underestimate mobilities even in disordered organic semiconductors, where even a little delocalisation can substantially enhance mobilities, as well as showing that three-dimensional calculations capture important delocalisation effects neglected in lower-dimensional approximations.
\end{abstract}

\maketitle

\begin{figure*}
    \centering
    \includegraphics[width=\textwidth]{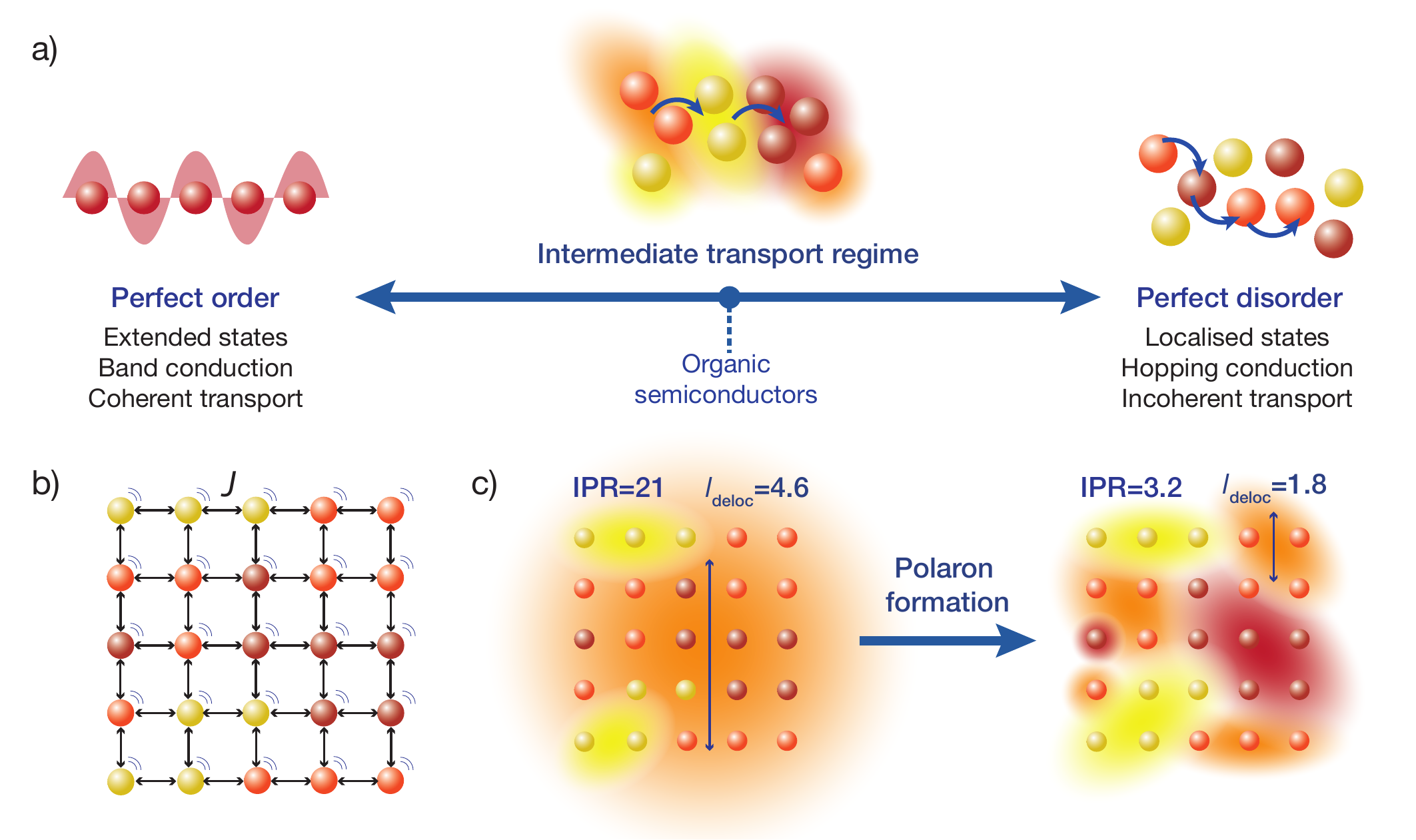}
    \caption{
    a) While the extreme regimes of transport, coherent band conduction through extended states and incoherent hopping between localised states, are well understood, materials such as organic semiconductors lie in the poorly understood intermediate regime. In this regime, charges hop between partially delocalised electronic states, where the rate of hopping depends on the overlap between the states. 
    b) We study a system that is a regular lattice of sites, with disordered energies (different colours), that are coupled to neighbours with an electronic coupling $J$ and to an environment (motion lines). 
    c) Delocalisation of the electronic states is found by diagonalising the system's Hamiltonian. Electronic couplings tend to delocalise the states, while disorder and environmental noise localise them. The resulting delocalisation of the electronic states can be quantified by the inverse participation ratio (IPR) (\cref{eq:IPR}) or the delocalisation length ($l_\mathrm{deloc}$) (\cref{eq:l_deloc}), which are included here for a material with $J=\SI{75}{meV}$ and $\sigma=\SI{150}{meV}$. When the formation of polarons is accounted for, the states become less delocalised.}
    \label{fig:regimes}
\end{figure*}

Charge and exciton transport is fundamental to materials science, particularly in applications for energy storage and conversion, including photovoltaics, batteries, light harvesting systems, lighting and electrocatalysts. However, many next-generation materials that promise significant functional improvements are disordered and noisy, making them difficult to treat mathematically and improve computationally. The clearest example of disordered electronic materials are organic semiconductors (OSCs)~\cite{kohler2015textbook}, but we expect that much of what we say here also applies to materials such as hybrid perovskites, conductive metal-organic frameworks and quantum dots. 

The difficulty with disordered materials is that they sit in the intermediate regime between the well-understood extremes of band conduction and hopping conduction (\cref{fig:regimes}a)~\cite{kohler2015textbook,Oberhofer2017}. In perfectly ordered crystals, charges move through Bloch waves, wavefunctions that are delocalised over infinitely many sites (individual atoms or molecules). By contrast, in extremely disordered materials (including some OSCs), electronic wavefunctions are localised to one molecule and charges move by thermally assisted hops to neighbouring sites. A theory of the intermediate transport regime must bridge these two qualitatively different extremes.

In disordered materials, two mechanisms localise electronic states away from infinite Bloch waves (\cref{fig:regimes}c). The first is Anderson localisation, which is caused by static disorder~\cite{Anderson1958}. The second is the formation of polarons due to a carrier's interaction with the environment~\cite{Frolich1954,Holstein1959,Grover1971}. Either mechanism, if strong enough, can localise states onto individual sites (giving so-called small polarons), but in the intermediate regime the localisation is not complete, and polaron states can be delocalised over multiple sites.

Many OSCs fall into the intermediate regime, which can be most clearly seen from failures of conventional simulations. Transport in most small-molecule OSCs (apart from organic crystals, where the importance of delocalisation has long been recognised~\cite{Grover1971}) is usually modelled as fully localised hopping, typically through a Gaussian density of states~\cite{Bassler1993}. The simplest hopping-rate expression is the Miller-Abrahams equation~\cite{Miller1960}, which neglects polaron formation, while Marcus theory accommodates both disorder and polaron formation~\cite{Marcus1956,Athanasopoulos2007,Fishchuk2013,Coropceanu}. These microscopic theories can be connected to measurable mobilities using kinetic Monte Carlo (KMC) simulations, a probabilistic approach based on averaging stochastic trajectories~\cite{Coropceanu}. KMC often significantly underestimates mobilities, requiring unphysically fast hopping rates (faster than molecular vibrations) to reproduce experimental results. 
This underestimation occurs because the assumption of completely localised states fails if inter-molecular couplings are comparable to the disorder or to the system-environment coupling, allowing polarons to delocalise across multiple sites~\cite{Bednarz2005,Oberhofer2012,Ferdinand2008,Yang2017,Liu2017,Rice2018}. An accurate description of intermediate-regime transport in OSCs is particularly important for understanding organic photovoltaics, where delocalisation has been proposed as the key explanation for how charges overcome their Coulomb attraction to achieve rapid and efficient charge separation~\cite{Tamura2013,Gelinas2012,Valleau2012,Bittner2014,Few2015,Huix-Rotllant2015,Bassler2015,Gluchowski2018,Jankovic2018,Athanasopoulos2019,Jankovic2020}.

Nevertheless, describing intermediate-regime transport has proven difficult, particularly in statically disordered systems~\cite{Oberhofer2017}. The challenge is that accounting for delocalisation requires a quantum-mechanical treatment, whose computational cost can balloon when disorder prevents periodic boundary conditions being used and forces large simulation boxes instead. In addition, mobility in disordered systems is often governed by deep traps, and long simulation times can be required to reach converged mobilities~\cite{Hoffmann2012}. 

Existing methods can be broadly divided into atomistic ones and those based on effective Hamiltonians. Atomistic calculations track the dynamics of both the nuclear degrees of freedom (usually using molecular mechanics) and the electronic ones (using quantum equations of motion)~\cite{Troisi2006,Fratini2016,Heck2016,Giannini2018,Giannini2019,Giannini2020,Ziogos2020}. Atomistic simulations do not have adjustable parameters, but they suffer from the considerable cost of tracking the atomic motion. As a result, taking OSCs as an example, the best atomistic simulations of charge transport are limited to about a thousand molecules, tracked for around \SI{1}{ps}~\cite{Giannini2018,Giannini2019,Giannini2020,Ziogos2020}. These capabilities enable remarkable simulations of layered organic crystals, which admit a two-dimensional simulation and are ordered enough that mobilities converge rapidly. However, the same approach cannot be applied to a three-dimensional disordered material that may require a nanosecond-long simulation. By contrast, effective-Hamiltonian models track fewer degrees of freedom~\cite{Savoie2014,Jackson2016,Jiang2016,Liu2017,Nematiaram2019,Sosorev2020,Varvelo2020}, allowing for larger and longer simulations. These approaches parametrise model Hamiltonians, which then govern the time evolution. Even if the parametrisation is accomplished using atomistic simulations, the important distinction is that an effective-Hamiltonian approach no longer tracks information about the individual atoms after the parametrisation. The weakness of these approaches is that they can neglect important phenomena if they are not included in the model Hamiltonian. For example, effective Hamiltonians designed to treat ordered organic crystals cannot treat disordered materials.

Several effective-Hamiltonian theories incorporate the three ingredients critical to describing the intermediate regime: delocalisation, disorder, and polaron formation~\cite{Bittner2014,Lee2015,Jankovic2020,Varvelo2020}. Most of these approaches describe polarons using the polaron transformation, which reduces the otherwise strong system-environment coupling, enabling a perturbative treatment of the remaining interactions. The differences between the approaches are in the details of the effective Hamiltonian and of the perturbative corrections, which give them strengths and weaknesses in different regimes. The Bittner-Silva theory~\cite{Bittner2014} is similar to what we do below, except that they study a bath of extended, shared phonons, an assumption that may not be appropriate in molecular systems where phonons are better described as local molecular vibrations~\cite{MayKuhn}. The Janković-Vukmirović approach~\cite{Jankovic2020} uses modified Redfield theory as its perturbation theory, making it valid when off-diagonal system-environment couplings are small. However, this condition is not guaranteed in some materials that interest us, including some OSCs; furthermore, disorder is the only localising feature in modified Redfield theory, meaning that the method cannot describe ordered systems where strong system-environment coupling results in small polarons~\cite{Jesenko2014}. The approach of Varvelo et~al.~\cite{Varvelo2020} differs from those above in that it uses a non-perturbative treatment of the system-environment couplings based on the hierarchy equations of motion. In principle, this technique is more accurate in the intermediate regime, but it is computationally more expensive and has so far only been used for one-dimensional chains.

We follow and extend the secular polaron-transformed Redfield equation (sPTRE)~\cite{Lee2015}, which has several key advantages. Most importantly, sPTRE is entirely in the polaron frame, which changes as a function of the system-environment coupling, allowing sPTRE to characterise intermediate-regime transport as well as exactly reproducing both the band-conduction and hopping-conduction extremes~\cite{Lee2015,Lee2012,Xu2016,Chang2013,Jang2008,Nazir2009,McCutcheon2010}. The up-front use of the polaron transformation also reduces the delocalisation of the electronic states~\cite{Rice2018} (\cref{fig:regimes}c), making mobility calculations easier. Its main limitation has been its computational cost, with sPTRE only ever applied to one-dimensional systems~\cite{Lee2015}.

Here, we overcome computational roadblocks that have limited sPTRE to one-dimensional systems to present the first three-dimensional description of partially delocalised carriers—whether charges or excitons—in intermediately disordered materials, over times as long as nanoseconds. Our results reproduce sPTRE in one dimension and hopping transport in the low-coupling limit, before showing that even small amounts of delocalisation can dramatically increase mobilities in two and, especially, three dimensions. We also show that these quantum-mechanical enhancements increase at low temperatures due to increasing polaron delocalisation.

\begin{figure*}
    \centering
    \includegraphics[width=\textwidth]{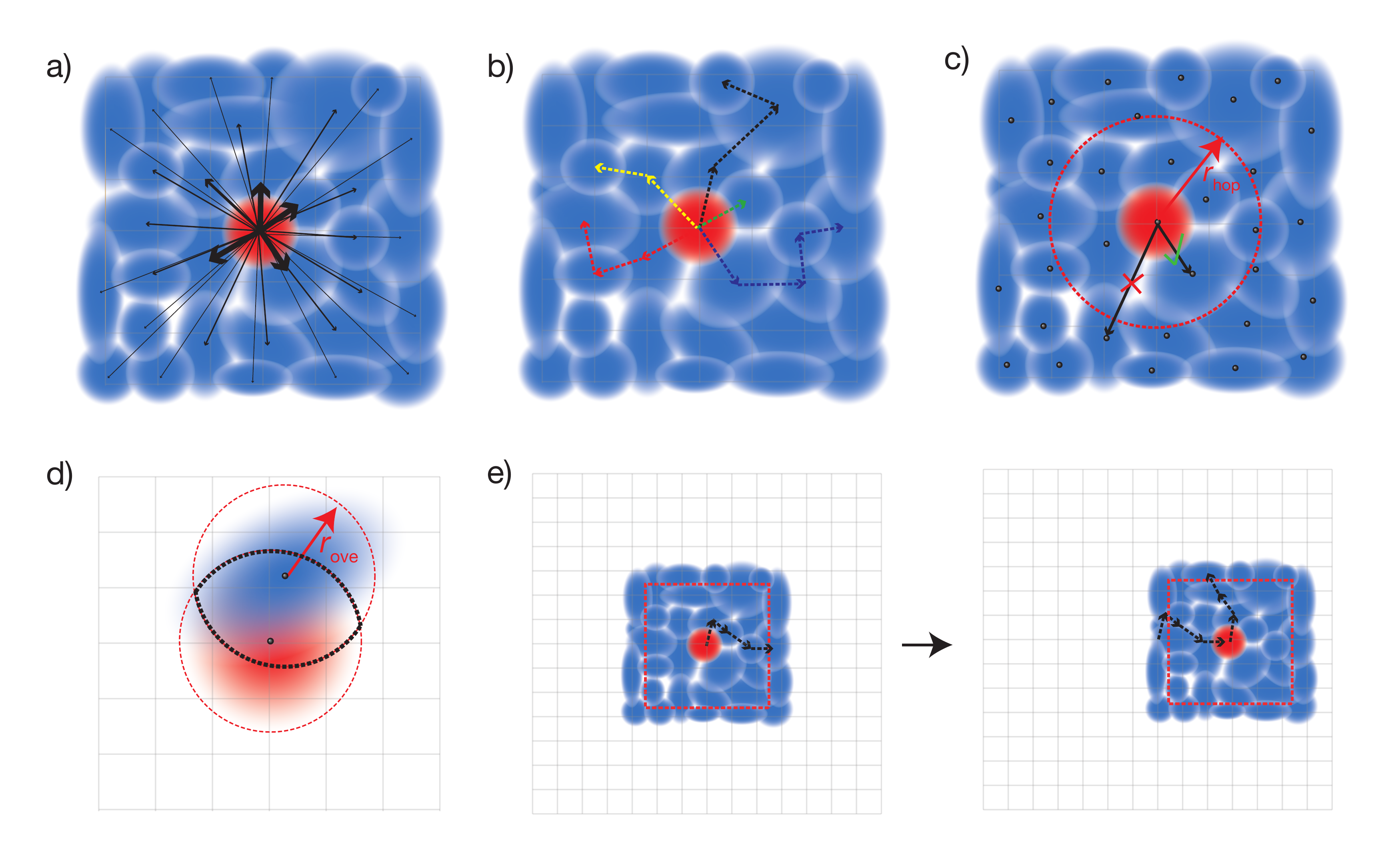}
    \caption{The four approximations underlying dKMC. 
    a) Full sPTRE master equation: the charge density can spread continuously throughout all polaron states. This approach is too expensive in more than one dimension. 
    b) Kinetic Monte Carlo: individual trajectories are formed from discrete, sequential hops, and are eventually averaged. 
    c) Hopping radius $r_\mathrm{hop}$: hops are only calculated for states whose centres (black dots) are close enough.
    d) Overlap radius $r_\mathrm{ove}$: only sites (grid points) that are close to both the initial and final polaron states are considered in calculating the hopping rate. 
    e) Diagonalising on the fly: instead of the whole Hamiltonian, only a subsystem of size $N_{box}^d$ is diagonalised at a time. As the charge moves too close to the boundary, a new Hamiltonian is re-diagonalised centred at the new location of the charge.}
    \label{fig:approximations}
\end{figure*}

\section{\label{sec:theory}Secular Polaron-Transformed Redfield Equation}

Our approach is based on sPTRE~\cite{Lee2015}, which we review in this section.

\paragraph{Hamiltonian}

We wish to describe an open quantum system, whose total Hamiltonian 
\begin{equation}
H_\mathrm{tot} = H_\mathrm{S} + H_\mathrm{B} + H_\mathrm{SB}
\end{equation}
consists of components describing the system ($H_\mathrm{S}$), the bath ($H_\mathrm{B}$) and the interaction between them ($H_\mathrm{SB}$).
All of the parameters introduced below that enter into $H_\mathrm{tot}$ could, in principle, be computed using atomistic simulations that combine molecular mechanics and quantum chemistry~\cite{kohler2015textbook,MayKuhn,Coropceanu,Few2015,Jankovic2020}.

Our system is a tight-binding model of a $d$-dimensional cubic lattice of $N^d$ sites, such as molecules or parts of molecules (\cref{fig:regimes}b). To represent disorder, the energy $E_n$ of each site $n$ is independently drawn from the Gaussian distribution $g(E)=\exp\left(-(E-E_0)^2/2\sigma^2\right)/\sqrt{2\pi\sigma^2},$ whose standard deviation $\sigma$ is the disorder of the material. The energetic disorder could arise from static variations in the orientation and spacing of molecules, producing a unique local environment around each molecule. The sites are assumed to be electronically coupled to nearest neighbours with coupling $J$, which enables delocalisation. We assume a constant nearest-neighbour coupling, although this assumption could easily be relaxed to allow off-diagonal disorder. Overall, the system Hamiltonian is therefore 
\begin{equation}
H_\mathrm{S} = \sum_n E_n \ket{n}\bra{n} + \sum_{m \neq n} J_{mn} \ket{m}\bra{n},
\end{equation}
where $\ket{n}$ represents the charge or exciton localised on site $n$.

We treat the environment as an independent, identical bath on every site, consisting of a series of harmonic oscillators, which can be thought of as vibrations of bonds in the molecules. The bath Hamiltonian is, therefore,
\begin{equation}
H_\mathrm{B}=\sum_{n,k}\omega_{nk} b^\dag_{nk}b_{nk},
\end{equation}
where the $k$th bath mode attached to the $n$th site has frequency $\omega_{nk}$, with creation and annihilation operators $b_{nk}$ and $b^\dag_{nk}$. Assuming a local bath is common in describing disordered molecular materials~\cite{kohler2015textbook,MayKuhn}; in crystalline systems, extended phonons that can couple different sites would be more appropriate.

The interaction between the system and the environment is treated by coupling every site to its bath with couplings $g_{nk}$, so that
\begin{equation}
H_\mathrm{SB} = \sum_{n,k} g_{nk}\ket{n}\bra{n}(b^\dag_{nk} + b_{nk}).
\end{equation}
This linear coupling model is a standard approximation, based on keeping the leading term in the Taylor expansion of a general system-bath interaction.

\paragraph{Polaron transformation}

Many materials have electronic ($J$) or system-bath couplings ($g_k$) that are too large to be treated as small perturbations. The polaron transformation reduces the system-environment coupling by absorbing it into the polaron itself, permitting the model to be treated using Redfield theory. Polaron formation is described using the state-dependent displacement operator~\cite{Grover1971}
\begin{equation}
e^S = e^{\sum_{n,k} \frac{g_{nk}}{\omega_{nk}}\ket{n}\bra{n}(b^\dag_{nk}-b_{nk})}.
\end{equation}
Applying it to the Hamiltonian incorporates lattice distortions into the system by displacing the environmental modes; the polaron-transformed Hamiltonian, indicated by tildes, is then 
\begin{equation}
\tilde{H}_\mathrm{tot} = e^S H_\mathrm{tot}e^{-S} = \tilde{H}_\mathrm{S} + \tilde{H}_\mathrm{B} + \tilde{H}_\mathrm{SB}.
\end{equation}
Here, the system Hamiltonian becomes
\begin{equation}
\tilde{H}_\mathrm{S} = \sum_n \tilde{E}_n \ket{n}\bra{n} + \sum_{m \neq n} J_{mn}\kappa_{mn} \ket{m}\bra{n},
\end{equation} 
where $\tilde{E}_n=E_n-\sum_k \abs{g_{nk}}^2/\omega_k$ and the coupling between sites is renormalised by the factor $\kappa_{mn}$,
\begin{equation}
\label{eq:kappa}
    \kappa_{mn} = e^{-\frac{1}{2}\sum_k\left[\frac{g^2_{mk}}{\omega^2_{mk}}\coth{\left(\frac{\beta\omega_{mk}}{2}\right)}+\frac{g^2_{nk}}{\omega^2_{nk}}\coth{\left(\frac{\beta\omega_{nk}}{2}\right)}\right]}.
\end{equation}
The bath Hamiltonian remains unchanged, $\tilde{H}_\mathrm{B} = H_\mathrm{B}$, while the system-bath Hamiltonian becomes
\begin{equation}
\tilde{H}_\mathrm{SB} = \sum_{n\ne m} J_{mn}\ket{m}\bra{n}V_{mn},
\end{equation}
with the new operator
\begin{equation}
V_{mn} = e^{\sum_k\frac{g_{mk}}{\omega_{mk}}\left(b^\dag_{mk}-b_{mk}\right)}e^{-\sum_k\frac{g_{nk}}{\omega_{nk}}\left(b^\dag_{nk}-b_{nk}\right)} - \kappa_{mn}.
\end{equation}

Summing many discrete vibrational modes is computationally costly, so we make two standard simplifications. First, we assume that system-bath couplings are identical at all sites, $g_{nk}=g_k$. Second, we assume that the spectral density $J(\omega)=\sum_kg_k^2\delta(\omega-\omega_k)$ is a continuous function~\cite{MayKuhn}.
The renormalisation factor then becomes
\begin{equation}
\kappa_{mn} = \kappa = e^{-\int_0^\infty \frac{d\omega}{\pi}\frac{J(\omega)}{\omega^2}\coth{\left(\beta\omega/2\right)}}.
\end{equation}
Because $\kappa<1$, the polaron transformation has two computational advantages: first, it reduces the electronic coupling $J$, making Redfield theory applicable, and second, it reduces the delocalisation of the electronic states obtained by diagonalising $\tilde{H}_S$~\cite{Rice2018} (\cref{fig:regimes}c).
Here, for concreteness, we adopt the widely used super-Ohmic spectral density $J(\omega) = \frac{\lambda}{2} (\omega/\omega_c)^3 \exp(-\omega/\omega_c)$,
where $\lambda$ is the reorganisation energy and $\omega_{c}$ is the cutoff frequency~\cite{Pollock2013,Jang2011,Jang2002,Wilner2015}. Unless specified otherwise, we use $\lambda=\SI{100}{meV}$, $\omega_c=\SI{62}{meV}$~\cite{Lee2015} and $T=\SI{300}{K}$. More generally, the equations above could be used for the more structured spectral densities of organic molecules.

\paragraph{Secular Redfield theory}

The polaron transformation reduces the system-bath coupling, allowing $\tilde{H}_\mathrm{SB}$ to be treated as a perturbation to the system~\cite{Lee2015}. Redfield theory is a second-order perturbative approach that, when applied in the polaron frame, results in the polaron-transformed Redfield equation (PTRE). It describes the evolution of the polaron-transformed reduced density matrix $\tilde{\rho}$ in the basis $\ket{\mu}$ of polaron states found by diagonalising $\tilde{H}_S$:
\begin{equation}
\label{eq:PTRE}
\frac{d\tilde{\rho}_{\mu\nu}(t)}{dt} = -i\omega_{\mu\nu}\tilde{\rho}_{\mu\nu}(t) + \sum_{\mu',\nu'}R_{\mu\nu,\mu'\nu'}\tilde{\rho}_{\mu'\nu'}(t),
\end{equation}
where $\omega_{\mu\nu}=E_\mu-E_\nu$. The Redfield tensor
\begin{multline}
R_{\mu\nu,\mu'\nu'} = \Gamma_{\nu'\nu,\mu\mu'} + \Gamma^*_{\mu'\mu,\nu\nu'} \\
- \delta_{\nu\nu'}\sum_\kappa\Gamma_{\mu\kappa,\kappa\mu'} - \delta_{\nu\nu'}\sum_\kappa \Gamma^*_{\nu\kappa,\kappa\nu'}
\end{multline}
describes the bath-induced relaxation in terms of damping rates
\begin{multline}
\label{eq:Gamma}
\Gamma_{\mu\nu,\mu'\nu'} = \sum_{m,n,m',n'} J_{mn} J_{m'n'} 
\\ \braket{\mu|m}\braket{n|\nu}\braket{\mu'|m'}\braket{n'|\nu'} K_{mn,m'n'}(\omega_{\nu'\mu'}),
\end{multline}
where
\begin{equation}
\label{eq:int_bath_cor}
K_{mn,m'n'}(\omega) = \int_0^\infty e^{i\omega \tau}\braket{\hat{V}_{mn}(\tau)\hat{V}_{m'n'}(0)}_{H_b}d\tau
\end{equation}
is the half-Fourier transform of the bath correlation function~\cite{Jang2011}
\begin{equation}
\braket{V_{mn}(\tau)V_{m'n'}(0)}_{H_b} = \kappa^2(e^{\lambda_{mn,m'n'}\phi(\tau)} - 1),
\end{equation}
where $\lambda_{mn,m'n'} = \delta_{mm'} - \delta_{mn'} + \delta_{nn'} - \delta_{nm'}$
and 
\begin{equation}
\phi(\tau) = \int_0^\infty \frac{d\omega}{\pi}\frac{J(\omega)}{\omega^2}\left[\cos(\omega \tau)\coth\left(\frac{\beta\omega}{2}\right) - i \sin(\omega \tau)\right].
\end{equation}

The PTRE of \cref{eq:PTRE} can be further simplified using the secular approximation to give the secular PTRE (sPTRE).
The polaron-transformed density matrix consists of diagonal populations and off-diagonal coherences. The evolution of $\tilde{\rho}_{\mu\nu}$ is controlled by the Redfield tensor, containing terms that transfer populations, dephase coherences, transfer coherences, and mix populations and coherences. In the interaction picture, these terms oscillate with a combined frequency of $\omega_{\nu\mu}-\omega_{\nu'\mu'}$. If this frequency is much greater than the inverse of the time frame $\Delta t$ over which the PTRE is solved, the oscillations are so rapid that the influence of these terms averages out to zero~\cite{MayKuhn}. The only terms that survive are those for which $\omega_{\nu\mu}-\omega_{\nu'\mu'} \ll \Delta t^{-1}$. This condition is met for population transfers and coherence dephasing, and the secular approximation is the assumption that only those terms survive. The result is sPTRE, in which populations and coherences are decoupled.

Furthermore, only populations are relevant for charge transport~\cite{Lee2015}, and they are invariant under the polaron transformation, $\rho_s(t)=\tilde{\rho}_s(t)$, leaving
\begin{equation}
\label{eq:diag_dens_evo}
\frac{d\rho_{\nu\nu}(t)}{dt} = \sum_{\nu'}R_{\nu\nu'}\rho_{\nu'\nu'}(t),
\end{equation}
where the Redfield tensor is now only two dimensional, containing only population transfer terms
\begin{equation}
\label{eq:secular_redfield_tensor}
R_{\nu\nu'} = 2\Re\left[\Gamma_{\nu'\nu,\nu\nu'}\right] - \delta_{\nu\nu'}\sum_\kappa2\Re\left[\Gamma_{\nu\kappa,\kappa\nu'}\right].
\end{equation}
The secular approximation reduces computational cost by reducing the number of density-matrix elements from $N^{2d}$ to $N^{d}$ and the number of Redfield-tensor elements from $N^{4d}$ to $N^{2d}$. 

The secular approximation does not significantly reduce the accuracy for the disordered systems we are studying, for two reasons. First, for parameters typical of the ones we survey, sPTRE agrees well~\cite{Lee2015} with a time-convolutionless second-order polaron master equation~\cite{Zimanyi2012} that does not use the secular or Markov approximations. Second, any inaccuracies will be minimal for long-time and long-range mobility calculations; at long times, the rate-limiting step in carrier diffusion is the thermal de-trapping from traps that have large energy differences from their neighbours.

\paragraph{Calculating mobilities}

The full time evolution of polaron-state populations is given by sPTRE (\cref{eq:diag_dens_evo}). \Cref{fig:approximations}a illustrates sPTRE evolution, where the charge density can spread to all other eigenstates in continuous time and in proportion to the corresponding Redfield rate. \Cref{eq:diag_dens_evo} has the solution
\begin{equation}
    \label{eq:sPTRE_solution}
    \rho(t)=\exp(Rt)\rho(0),
\end{equation}
which can be used to calculate the expectation value of the mean-squared displacement of the charge, $\langle r^2(t)\rangle = \mathrm{tr}(r^2\rho(t))$, at any time. 

To calculate the mobility, $\langle r^2(t)\rangle$ is averaged over many realisations of disorder ($n_\mathrm{iters}$), i.e., disordered energy landscapes generated using the same microscopic parameters. The resulting average $\overline{\langle r^2(t)\rangle}$ determines the diffusion constant
\begin{equation}
D=\lim_{t\to\infty}\frac{d}{dt}\left(\frac{\overline{\langle r^2(t)\rangle}}{2d}\right).    
\end{equation}
Finally, for a carrier of charge $q$, the mobility is given by the Einstein relation
\begin{equation}
    \label{eq:einstein_mobility}
    \mu=\frac{qD}{k_BT}.
\end{equation}

\begin{figure*}
    \fbox{
    \begin{minipage}{0.97\textwidth}
        \setlist{nolistsep}
        \begin{minipage}[t]{0.49\textwidth}\vspace{0pt}%
            \raggedright
            Steps to be carried out for every set of microscopic parameters $N$, $d$, $\sigma$, $J$, $\lambda$, $\omega_c$ and $T$:
            \begin{enumerate}[leftmargin=*]
                \item (Calibrating cutoff radii) For $n_\mathrm{calib}$ realisations of disorder:
                \begin{enumerate}[leftmargin=*,label=\alph*.]
                    \item Generate an $N^d$ lattice of random energies.
                    \item Set $r_\mathrm{hop}\leftarrow 0$ and $r_\mathrm{ove}\leftarrow 0$.
                    \item While $T_{r_\mathrm{hop}-1,r_\mathrm{ove}}/T_{r_\mathrm{hop},r_\mathrm{ove}}<a_\mathrm{hop}$:
                    \begin{enumerate}[leftmargin=*]
                        \item Update $r_\mathrm{hop}\leftarrow r_\mathrm{hop}+1$.
                        \item While $T_{r_\mathrm{hop},r_\mathrm{ove}-1}/T_{r_\mathrm{hop},r_\mathrm{ove}}<a_\mathrm{ove}$:
                        \begin{enumerate}[leftmargin=*]
                            \item Update $r_\mathrm{ove}\leftarrow r_\mathrm{ove}+1$.
                            \item Update $N_\mathrm{box} \leftarrow 2(r_\mathrm{ove}+r_\mathrm{hop})$.
                            \item Create a polaron-transformed Hamiltonian $\tilde{H}_S$ of size $N_\mathrm{box}^d$ at the centre of the lattice and find the polaron states, their centres and their energies.
                            \item Choose polaron state $\nu$ closest to the centre of the lattice.
                            \item Create a list $L$ of all polaron states $\nu'$ such that $\abs{\vec{C}_\nu-\vec{C}_{\nu'}} < r_\mathrm{hop}$.
                            \item Calculate $R_{\nu\nu'}$ for all $\nu'\in L$ using \cref{eq:secular_redfield_tensor}, only summing in \cref{eq:Gamma} over sites $\vec{q}$ such that $\abs{\vec{q}-\vec{C}_{\nu}} < r_\mathrm{ove}$ and $\abs{\vec{q}-\vec{C}_{\nu'}} < r_\mathrm{ove}$. 
                            \item Set $T_{r_\mathrm{hop},r_\mathrm{ove}} \leftarrow \sum_{\nu'\in L}R_{\nu\nu'}$.
                        \end{enumerate}
                    \item Update $r_\mathrm{ove}\leftarrow r_\mathrm{ove}-1$.
                    \end{enumerate}
                    \item Update $r_\mathrm{hop}\leftarrow r_\mathrm{hop}-1$.
                \end{enumerate}
                \item Average $r_\mathrm{hop}$ and $r_\mathrm{ove}$ over the $n_\mathrm{calib}$ realisations and set $N_\mathrm{box}\leftarrow 2(r_\mathrm{hop}+r_\mathrm{ove})$.
            \end{enumerate}
        \end{minipage}
        \hfill
        \begin{minipage}[t]{0.49\textwidth}\vspace{0pt}%
            \begin{enumerate}[leftmargin=*,start=3]
                \item (Kinetic Monte Carlo) For $n_\mathrm{iter}$ realisations of disorder:
                \begin{enumerate}[leftmargin=*, label=\alph*.]
                    \item Generate an $N^d$ lattice of random energies.
                    \item For $n_\mathrm{traj}$ trajectories:
                    \begin{enumerate}[leftmargin=*]
                        \item Create polaron-transformed Hamiltonian $\tilde{H}_S$ of size $N_\mathrm{box}^d$ at the centre of the lattice and find the polaron states, their centres and their energies.
                        \item Set $t\leftarrow 0$ and choose initial polaron state $\nu$ closest to the centre of the lattice.
                        \item Repeat until $t>t_\mathrm{end}$:
                        \begin{enumerate}[leftmargin=*]
                            \item Create a list $L$ of all polaron states $\nu'$ such that  $\abs{\vec{C}_\nu-\vec{C}_{\nu'}} < r_\mathrm{hop}$.
                            \item Calculate $R_{\nu\nu'}$ for all $\nu'\in L$ using \cref{eq:secular_redfield_tensor}, only summing in \cref{eq:Gamma} over sites $\vec{q}$ such that $\abs{\vec{q}-\vec{C}_{\nu}} < r_\mathrm{ove}$ and $\abs{\vec{q}-\vec{C}_{\nu'}} < r_\mathrm{ove}$.
                            \item Set $S_{\nu'}\leftarrow\sum_{\mu=1}^{\nu'} R_{\nu\mu}$ for all $\nu'\in L$. 
                            \item Set $T\leftarrow \sum_{\nu'\in L}S_{\nu'}$.
                            \item Select the destination state by finding $\nu'$ such that $S_{\nu'-1} < uT < S_{\nu'}$, for uniform random number $u \in (0,1]$, and update $\nu\leftarrow\nu'$. 
                            \item Update $t\leftarrow t+\Delta t$, where $\Delta t = -T^{-1}\ln{v}$ for uniform random number $v \in (0,1]$.
                            \item If the charge is within $r_\mathrm{hop}+r_\mathrm{ove}$ of the edge of the diagonalised Hamiltonian, diagonalise a new Hamiltonian of size $N_\mathrm{box}^d$ centred on the charge. 
                        \end{enumerate}
                    \end{enumerate}
                \end{enumerate}
                \item Calculate $\mu$ using \cref{eq:einstein_mobility}. 
            \end{enumerate}
        \end{minipage}
    \end{minipage}
    }
    \caption{The delocalised kinetic Monte Carlo algorithm.}
    \label{listing}
\end{figure*}

\section{\label{sec:3D} Delocalised Kinetic Monte Carlo}

sPTRE has only been applied to one-dimensional systems~\cite{Lee2015}, because of three computational hurdles. First, generating the polaron states by diagonalising the $N^{d}$ $\tilde{H}_\mathrm{S}$ scales as $O(N^{3d})$, where $N$ is the number of sites along each of the $d$ dimensions. Second, tracking the population transfer between all pairs of polaron states involves calculating the full Redfield tensor $R_{\nu\nu'}$ (\cref{eq:secular_redfield_tensor}), which has $N^{2d}$ elements. Lastly, each population transfer rate depends on the damping rates $\Gamma$ (\cref{eq:Gamma}), calculating which involves a sum over $N^{4d}$ sites to account for the spatial overlap of the polaron states. Therefore, sPTRE scales as $O(N^{3d}) + O(N^{6d})$ overall, which, for reasonably sized lattices, is manageable only for $d=1$. 

Our approach, dKMC, overcomes these limitations using four approximations (\cref{fig:approximations}):
\begin{enumerate}
    \item \textbf{Kinetic Monte Carlo:} We reduce the number of Redfield rates that need to be calculated by mapping sPTRE onto kinetic Monte Carlo (KMC). Rather than tracking the time-dependent populations of all polaron states, we track stochastic trajectories through the polaron states (\cref{fig:approximations}b), followed by averaging. This approach mirrors standard KMC, which also probabilistically integrates a large master equation. Individual trajectories are found by hopping to another polaron state with probability proportional to the corresponding Redfield rate for population transfer. Hopping continues until a pre-determined end time $t_\mathrm{end}$. Therefore, instead of calculating Redfield rates for every pair of polaron states, as in sPTRE, KMC requires only calculate outgoing rates at each step. This reduces the number of Redfield rates that need to be calculated from $N^{2d}$ to $N^{d}n_\mathrm{hop}n_\mathrm{traj}$, where $n_\mathrm{hop}$ is the number of hops, which depends on $t_\mathrm{end}$, and $n_\mathrm{traj}$ is the number of trajectories, which controls the final averaging error.
    
    To calculate $\langle r^2(t)\rangle$ for \cref{eq:einstein_mobility}, we assume the charge occupying a particular polaron state (at a particular time) is located at its centre, defined as the expectation value of the position, $\vec{C}_\nu=\bra{\nu}\vec{r}\ket{\nu}$; $\langle r^2(t)\rangle$ is then the average of the square of this displacement over all the trajectories.
    
    \item \textbf{Hopping cutoff radius:} We reduce the number of Redfield rates to be calculated by introducing a hopping cutoff radius, $r_\mathrm{hop}$. For polaron states that are far away from the current state, the spatial overlaps, and therefore Redfield rates, are very small compared to polaron states that are close by. Therefore, we only calculate rates to polaron states whose centre $\vec{C}_\nu$ lies within $r_\mathrm{hop}$ of the centre of the current state (\cref{fig:approximations}c). The error in this approximation is tunable, because $r_\mathrm{hop}$ can can be arbitrarily increased depending on the desired accuracy. We choose our $r_\mathrm{hop}$ by gradually increasing it by one lattice spacing until the total sum of outgoing rates to states with centres within $r_\mathrm{hop}$ converges, not changing by more than a target factor $a_\mathrm{hop}$ between increments. The hopping cutoff radius reduces the number of Redfield rates to be calculated at each hop from $N^d$ to $O(r_\mathrm{hop}^d)$, thus reducing the total number of rates that need to be calculated in each random energetic landscape to $r_\mathrm{hop}^d n_\mathrm{hop}n_\mathrm{traj}$.
    
    \item \textbf{Overlap cutoff radius:} We reduce the cost of calculating individual Redfield rates by introducing an overlap cutoff radius, $r_\mathrm{ove}$. While eigenstates do, in principle, spread across the entire lattice, Anderson localisation predicts that their amplitude decreases exponentially with distance from their centre. Therefore, in calculating damping rates (\cref{eq:Gamma}), we only sum over sites that are simultaneously within a distance of $r_\mathrm{ove}$ from the centres of both polaron states (\cref{fig:approximations}d). Again, the error in this approximation is tunable, and can be decreased arbitrarily by increasing $r_\mathrm{ove}$. 
    We choose our $r_\mathrm{ove}$ by gradually increasing it by one lattice spacing until the sum of outgoing Redfield rates calculated by only including sites within $r_\mathrm{ove}$ of both the initial and final states converges, not changing by more than a target factor $a_\mathrm{ove}$ between increments. The overlap cutoff radius reduces the number of sites summed over in each damping rate from $N^{4d}$ to $O(r_\mathrm{ove}^{4d})$.

    In practice, we calculate $r_\mathrm{hop}$ and $r_\mathrm{ove}$ simultaneously, as outlined in \cref{listing}, by progressively increasing both until the total sum of the outgoing Redfield rates converges onto a desired accuracy. For simplicity, we choose the two target accuracies to be equal, $a_\mathrm{dKMC}=a_\mathrm{hop}=a_\mathrm{ove}$.
    
    \item \textbf{Diagonalising on the fly:} We reduce the time required to calculate the polaron states by diagonalising the Hamiltonian on the fly. Rather than diagonalise the entire lattice, we only diagonalise a subset of the Hamiltonian of size $N_\mathrm{box}^d$ centred at the location of the charge (\cref{fig:approximations}e). The charge moves within the box until it gets too close to the edge, which we set to be within $r_\mathrm{hop}+r_\mathrm{ove}$ of the edge. This buffer ensures we can accurately describe the next hop, which requires a distance of $r_\mathrm{hop}$ to contain the centres of all relevant polaron states, as well as a further distance of $r_\mathrm{ove}$ to ensure the entirety of polaron states at the edge of $r_\mathrm{hop}$ are well defined. In three dimensions, diagonalisation is usually the computational bottleneck, so we make the box as small as possible, $N_\mathrm{box}=2(r_\mathrm{hop}+r_\mathrm{ove})$. Once the charge leaves the buffer, the Hamiltonian corresponding to a new box of size $N_\mathrm{box}^d$ centred at the new location of the charge is re-diagonalised. The landscape continues to be updated as the charge hops through the material, ultimately reducing the cost of calculating the polaron states from $O(N^{3d})$ to $O(N_\mathrm{box}^{3d}n_{\mathrm{hop}})$.
\end{enumerate}

Overall, the four approximations above transform sPTRE to dKMC and make it possible to model three-dimensional charge transport in disordered, noisy materials. The detailed steps involved in the algorithm are shown in \cref{listing}. Overall, the scaling of the technique has been reduced from sPTRE's $O(N^{3d}) + O(N^{6d})$ to $O(N_\mathrm{box}^{3d}n_\mathrm{hop}) + O(r_\mathrm{ove}^{4d} r_\mathrm{hop}^d n_\mathrm{hop} n_\mathrm{traj})$ for dKMC. For example, for a disordered 3D system with $J/\sigma=0.1$ and $N=100$, the scaling is reduced by at least 25 orders of magnitude.

\section{\label{sec:results}Results and Discussion}

\paragraph{Accuracy}

\begin{figure}
    \centering
    \includegraphics[width=\columnwidth]{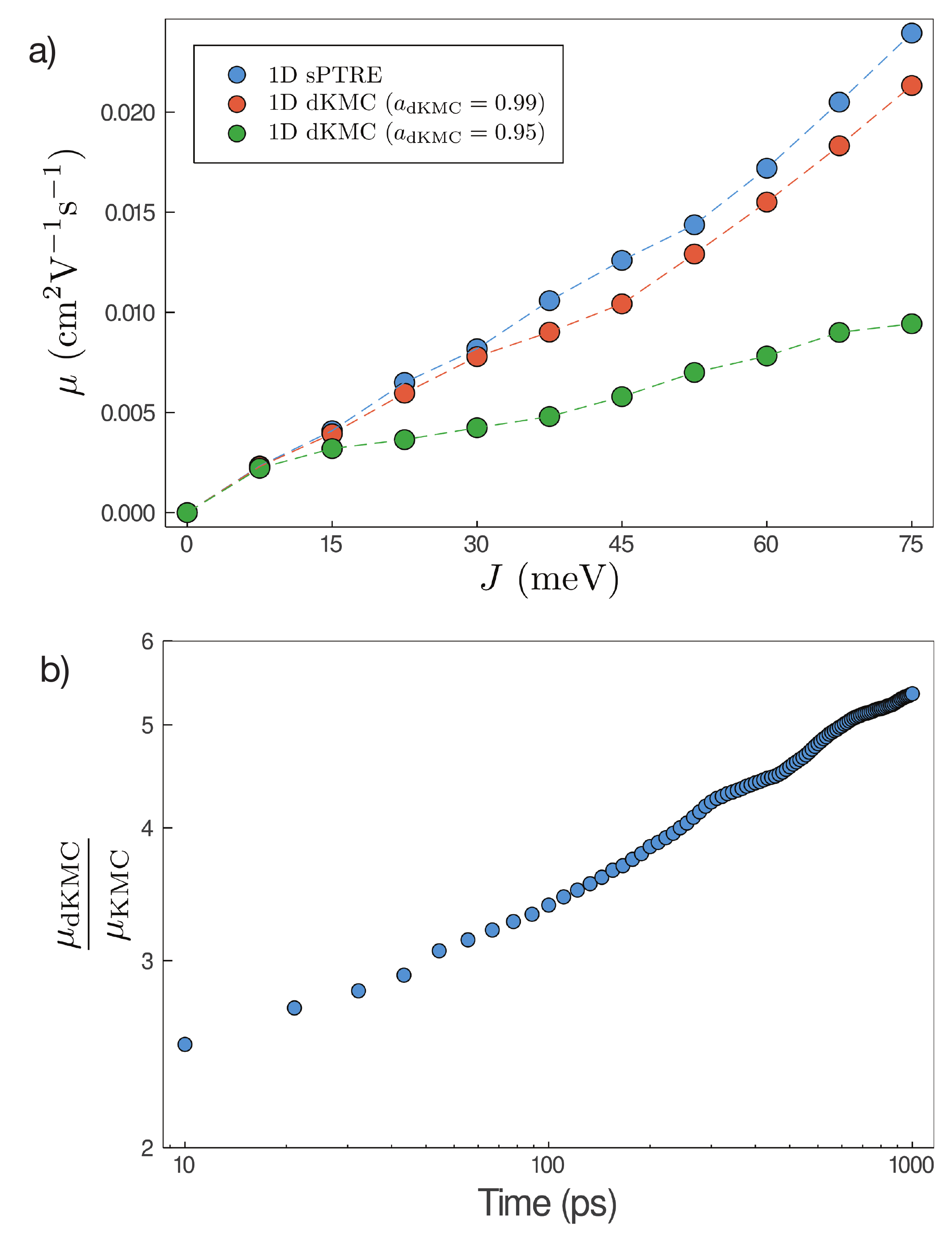}
    \caption{a) Mobilities ($\mu$) in one dimension at \SI{100}{ps} predicted by both sPTRE and dKMC, with disorder $\sigma=\SI{150}{meV}$ and shown as a function of the electronic coupling $J$. dKMC can be made arbitrarily accurate by increasing the accuracy parameter $a_\mathrm{dKMC}$ ($a_\mathrm{dKMC}=0.99$ is used elsewhere in this paper).
    b) Throughout this paper, mobilities—which are time-dependent~\cite{Hoffmann2012}—are calculated at \SI{100}{ps}. The mobility enhancement due to delocalisation ($\mu_\mathrm{dKMC}/\mu_\mathrm{KMC}$) increases with the time (here shown for one dimension), meaning that longer-time delocalisation enhancements would always be at least as large as reported here.}
    \label{fig:benchmarking}
\end{figure}

All of the approximations in dKMC are controllable, meaning that the error can be arbitrarily reduced given additional computational resources. This accuracy can be demonstrated by comparing dKMC mobilities to those predicted by sPTRE in one dimension, and \cref{fig:benchmarking}a shows the agreement increasing with the accuracy parameter $a_\mathrm{dKMC}$. The agreement between sPTRE and dKMC at $a_\mathrm{dKMC}=0.99$ leads us to adopt that value throughout this paper. Cutoff radii always lead to an underestimation of delocalisation effects, meaning that the substantial delocalisation enhancements reported below are strictly lower limits.

It is also necessary to choose the time cutoff $t_\mathrm{end}$, because mobilities in disordered materials are time dependent (or dispersive), and it can take a long time to converge on a steady-state mobility~\cite{Hoffmann2012}. The ultimate choice will depend on particular applications; we report mobilities at $t_\mathrm{end} = \SI{100}{ps}$, which corresponds to charge transit times on typical length scales in OSCs (tens of nanometres). For our purposes, the important fact is that mobility enhancements due to delocalisation at longer times are always at least as large as at $t_\mathrm{end}$ (\cref{fig:benchmarking}b). At longer times, charges are increasingly likely to get stuck in deeper traps, causing the mobility to decrease with time. Delocalisation allows the wavefunction to leak onto neighbouring sites, helping the detrapping and giving larger enhancements at longer times.

\paragraph{Importance of 3D effects}

\begin{figure}
    \centering
    \includegraphics[width=\columnwidth]{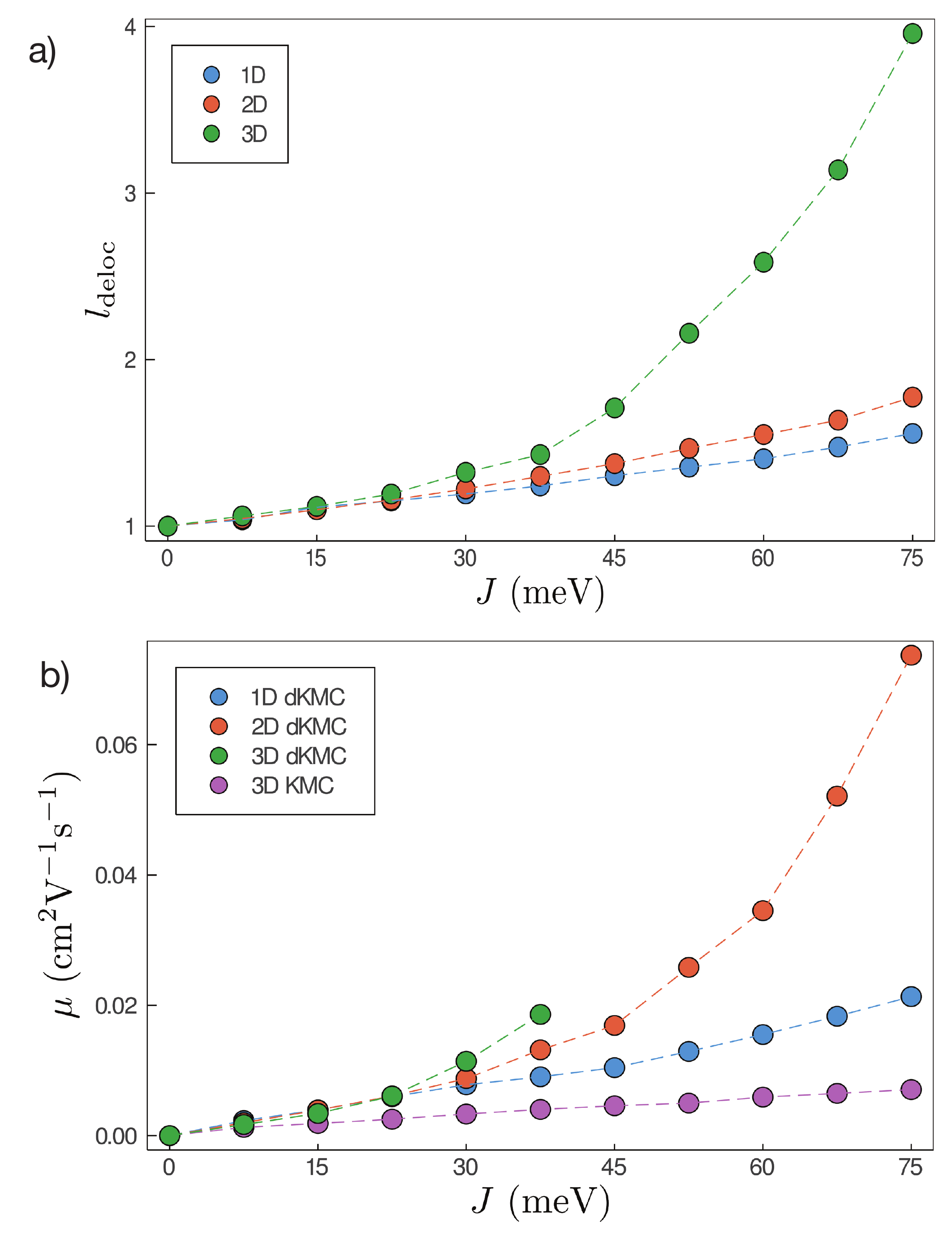}
    \caption{a) Delocalisation length ($l_\mathrm{deloc}$) in each dimension as a function of the electronic coupling ($J$), for disorder $\sigma=\SI{150}{meV}$. All else being equal, the delocalisation length is considerably greater in three dimensions than in lower dimensions. 
    b) Mobilities ($\mu$) at 100 ps calculated by dKMC and standard KMC as a function of $J$, with $\sigma=\SI{150}{meV}$. We only include KMC in three dimensions for legibility, because KMC mobilities in the lower dimensions are similar in magnitude. KMC and dKMC agree in the low-coupling limit when states are localised, but as the electronic coupling and delocalisation increase, so too do the mobilities predicted by dKMC compared to those predicted by KMC.}
    \label{fig:mobility_enhancement}
\end{figure}

dKMC captures effects missing in lower-dimensional approximations, in particular the extent of delocalisation of the polaron states.  We define the delocalisation length
\begin{equation}
    \label{eq:l_deloc}
    l_\mathrm{deloc} = \sqrt[d]{\overline{\mathrm{IPR}_\nu}},
\end{equation}
where we average the inverse participation ratios of the polaron states, 
\begin{equation}
    \label{eq:IPR}
    \mathrm{IPR}_\nu = \frac{1}{\sum_n \abs{\braket{n|\nu}}^4},
\end{equation}
which indicates the number of sites $n$ that a state $\nu$ extends over~\cite{Moix2013}. Therefore, $l_\mathrm{deloc}$ describes the size of a state along each dimension, and is a way of comparing the extent of delocalisation across different dimensions.

\Cref{fig:mobility_enhancement}a shows that $l_\mathrm{deloc}$ increases as a function of $J$ in all three dimensions, as expected. More interestingly, $l_\mathrm{deloc}$ is significantly larger in three dimensions than in one or two, with the difference becoming larger with increasing $J$. Therefore, including all three dimensions is essential for modelling intermediately delocalised charge transport, and lower-dimensional models may significantly underestimate delocalisation effects.

\paragraph{Delocalisation enhances mobility}

Our most important finding is that even modest delocalisation dramatically enhances mobilities, meaning that delocalisation is critical for explaining transport in the intermediate regime. As $J$ increases, the increase in delocalisation increases overlaps between states and, therefore, the Redfield transfer rates and the ultimate mobilities (\cref{fig:mobility_enhancement}b). For values of $J$ and $\sigma$ that are reasonable for OSCs, including delocalisation using dKMC can increase mobilities by close to an order of magnitude above the localised-hopping of standard KMC, which helps explain why mobilities predicted by KMC are usually too low compared to experiment. Furthermore, these enhancements require only a small amount of delocalisation, less than 2 sites in each direction in 2D.

\paragraph{Larger enhancements at lower temperatures}

\begin{figure}
    \centering
    \includegraphics[width=\columnwidth]{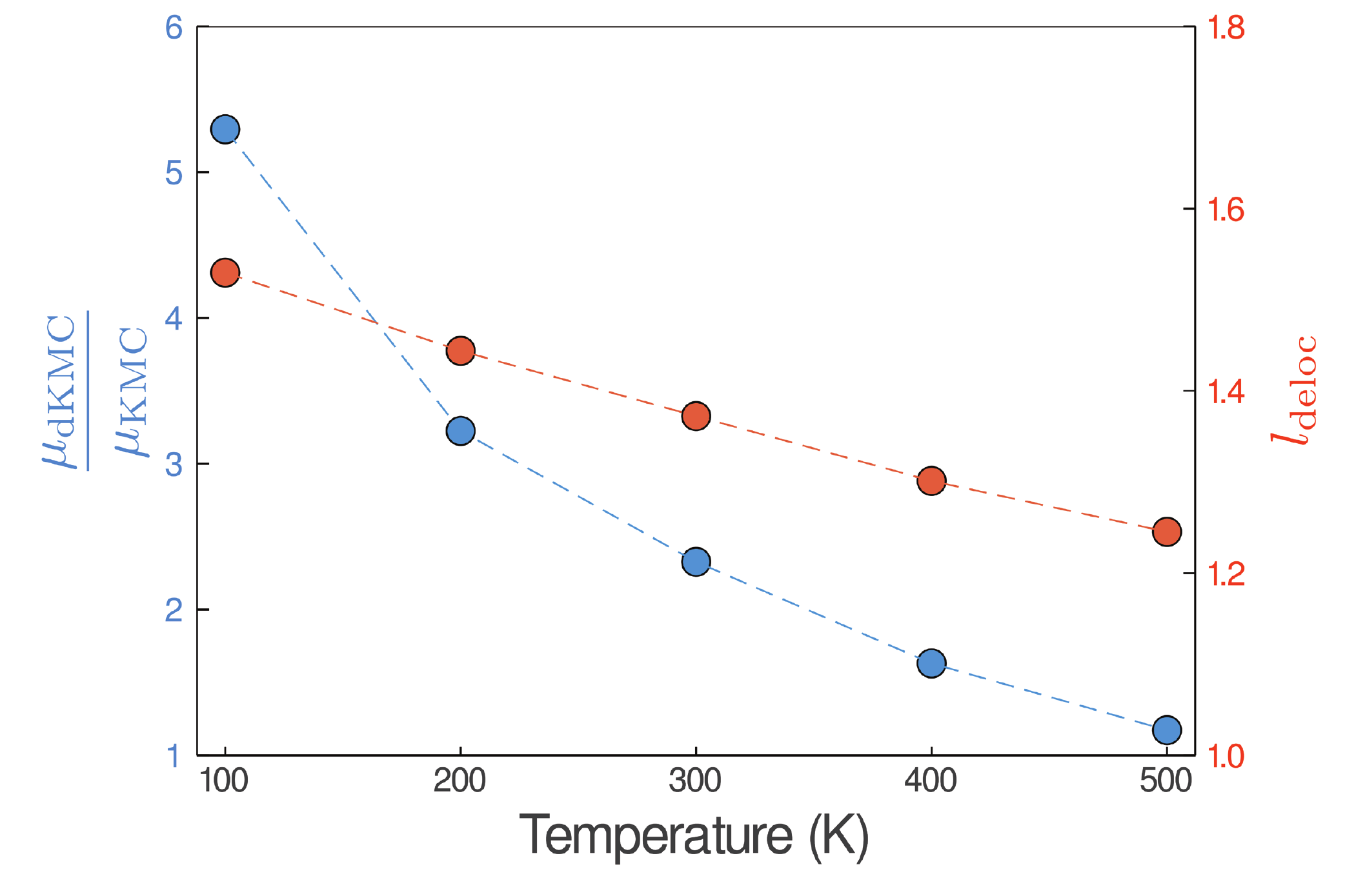}
    \label{fig:temperature}
    \caption{The mobility enhancement ($\mu_\mathrm{dKMC}/\mu_\mathrm{KMC}$), shown here in two dimensions for $J=\SI{45}{meV}$ and $\sigma=\SI{150}{meV}$, increases as the temperature is reduced, due to the increase in delocalisation of the polaron states ($l_\mathrm{deloc}$).}
\end{figure}

Finally, the delocalisation mobility enhancement is temperature dependent, with larger enhancements at low temperatures (\cref{fig:temperature}). The higher mobilities at lower temperatures are caused by the increased delocalisation of the polaron states. Applying the polaron transformation reduces $J$ by multiplying by a factor of $\kappa$ (\cref{eq:kappa}), where $\kappa<1$. $\kappa$ increases as $T$ is lowered and, therefore, $J$ is reduced less at low $T$ than at higher $T$, allowing the polaron states to delocalise further and assisting their mobility.

\paragraph{Outlook}

The most immediate area for future work is the application of the newly developed theory and the important trends identified in this paper to the prediction of experimental mobilities in concrete disordered materials. As we have shown, standard KMC approaches underestimate mobilities; therefore, we expect that including delocalisation will improve experimental agreement. However, connecting dKMC to experimental mobilities still faces some challenges, especially the difficulty of obtaining good estimates of disorder, whether experimentally~\cite{Hood2019} or using \textit{ab initio} calculations.

The main limitation of dKMC remains computational cost; in particular, in \cref{fig:mobility_enhancement}b, 3D dKMC is limited to modest values of $J$. As the states become larger, the Hamiltonian box needs to be increased to accurately capture the states and allow them to move around. The large box is expensive to diagonalise, and it requires the calculation of more rates, with more sites contributing to the overlap calculations in each rate. Nevertheless, there is a clear trend, and we expect the importance of delocalisation to be even more pronounced at high $J$ in 3D. In the future, it may be possible to reduce the computational cost further, either through additional approximations, or by identifying robust trends in the numerical results that can then be extrapolated.

dKMC is also limited by the approximations made in sPTRE, some of which could be relaxed using other master equations. sPTRE makes the Markov and secular approximations, and it can be inaccurate for systems that are weakly coupled to slow environments~\cite{Lee2012,Pollock2013,Lee2015}. The parameter regimes we studied fall within the range of validity of sPTRE established in previous work~\cite{Lee2015}. However, extending dKMC to weakly coupled slow baths would require modifications to the underlying sPTRE, for example by incorporating a variational polaron transformation~\cite{Silbey:1984,Zimanyi2012,Pollock2013} instead of the fully displaced version used in sPTRE. The variational treatment would also enable the treatment of Ohmic and sub-Ohmic baths, unlike the super-Ohmic ones assumed in sPTRE.

Furthermore, we expect that it will be possible to extend dKMC to other commonly encountered situations. The most straightforward extensions would include the prediction of mobilities at high charge densities, in the presence of external electric fields, on irregular or anisotropic lattices, or in spatially constrained domains. It may also be possible to extend dKMC to describe the more difficult problem of charge separation of excitons in organic photovoltaics. Because charge separation is a two-body problem involving the correlated motion of an electron and a hole, the computational difficulty is roughly the square of the single-body mobility calculation, meaning that a fully quantum-mechanical treatment has so far proved intractable in three dimensions~\cite{Few2015}. We expect that dKMC will make this problem computationally accessible, allowing the first simulation of the full dynamics (and, therefore, efficiency) of charge separation in the presence of disorder, delocalisation, and noise. A complete kinetic model would help settle the debate about the main drivers of charge separation, and unite the proposed mechanisms including delocalisation~\cite{Gelinas2012}, entropy~\cite{Clarke2010,Hood2016} and energy gradients~\cite{Jamieson2012}. 

\section{Conclusions}

dKMC is the first approach able to describe charge transport in intermediately disordered materials in three dimensions. It keeps the benefits of sPTRE—fully quantum dynamics, accurate treatment of polarons, and the ability to reproduce both extremes of transport—while overcoming computational obstacles that have prevented sPTRE from being used in more than one dimension. We have used dKMC to capture the effects of delocalisation and show that carrier mobilities are significantly higher than those predicted by standard KMC. Indeed, even small amounts of delocalisation—less than two sites—can increase mobilities by an order of magnitude. All of these mobility enhancements are greater at lower temperatures, due to the increased delocalisation of polaron states. In the future, we expect that dKMC can be extended to a wider range of systems, shedding even more insight into fundamental charge- and exciton-transport processes.

\begin{acknowledgments}
We were supported by a Westpac Scholars Trust Research Fellowship, a Westpac Scholars Trust Future Leaders Scholarship, an Australian Government Research Training Program scholarship, and a University of Sydney Nano Institute Grand Challenge.
We were supported by computational resources and assistance from the National Computational Infrastructure (NCI), which is supported by the Australian Government, and by the Sydney Informatics Hub and the University of Sydney's high-performance computing cluster Artemis.
\end{acknowledgments}

\end{document}